\documentclass[]{emulateapj}

\usepackage{graphics,graphicx,xspace,natbib,amssymb}
\usepackage{amsmath}
\usepackage{threeparttable}

\newcommand{\psbs}{post-starburst galaxies\xspace}

\newcommand{\comment}[1]{}
\newcommand{\colorgrads}{color gradients\xspace}
\newcommand{\colorgrad}{color gradient\xspace}
\newcommand{\cgstrength}{color gradient strength\xspace}

\accepted{August 6, 2020}

\shorttitle{Color Gradients along the Quiescent Sequence}
\shortauthors{Suess et al.}

\begin{document}

\title{color gradients along the quiescent galaxy sequence:\\clues to quenching and structural growth}

\author{Katherine A. Suess\altaffilmark{1}, Mariska Kriek\altaffilmark{1}, Sedona H. Price\altaffilmark{2}, Guillermo Barro\altaffilmark{3}} 

\altaffiltext{1}{Astronomy Department, University of California, Berkeley, CA 94720, USA}
\altaffiltext{2}{Max-Planck-Institut f{\"u}r extraterrestrische Physik, Postfach 1312, Garching, 85741, Germany}
\altaffiltext{3}{Department of Physics, University of the Pacific, 3601 Pacific Ave, Stockton, CA 95211, USA}
\email{suess@berkeley.edu}

\begin{abstract}
This Letter examines how the sizes, structures, and \colorgrads of galaxies change along the quiescent sequence. Our sample consists of $\sim$400 quiescent galaxies at $1.0\le z\le2.5$ and $10.1 \le \log{M_*/M_\odot}\le11.6$ in three CANDELS fields. We exploit deep multi-band {\it HST} imaging to derive accurate mass profiles and \colorgrads, then use an empirical calibration from rest-frame $UVJ$ colors to estimate galaxy ages. We find that--- contrary to previous results--- the youngest quiescent galaxies are {\it not} significantly smaller than older quiescent galaxies at fixed stellar mass. These `post-starburst' galaxies only appear smaller in half-light radii because they have systematically flatter \colorgrads. The strength of \colorgrads in quiescent galaxies is a clear function of age, with older galaxies exhibiting stronger negative \colorgrads (i.e., redder centers). Furthermore, we find that the central mass surface density $\Sigma_1$ is independent of age at fixed stellar mass, and only weakly depends on redshift. This finding implies that the central mass profiles of quiescent galaxies do not significantly change with age; however, we find that older quiescent galaxies have additional mass at large radii. Our results support the idea that building a massive core is a necessary requirement for quenching beyond $z=1$, and indicate that \psbs are the result of a rapid quenching process that requires structural change. Furthermore, our observed color gradient and mass profile evolution supports a scenario where quiescent galaxies grow inside-out via minor mergers.
\end{abstract}


%------------------------------------------------------------------------------------------------------------------------------------------------------------------------
\section{Introduction}

Building a dense central core appears to be a prerequisite for forming massive quiescent galaxies \citep[e.g.][]{cheung12,fang13,vandokkum14,tacchella15b,whitaker17,mosleh17}. Galaxies cease forming stars after reaching a threshold in central mass surface density $\Sigma_1$ \citep[e.g.][]{barro18,lee18,woo19} or central velocity dispersion \citep[e.g.][]{vandokkum15}. At the same time, the observed variations in the properties of quiescent galaxies across cosmic time indicate that  galaxies continue to evolve after quenching their star formation, likely due to growth via dry minor mergers \citep[e.g.,][]{bezanson09,naab09,hopkins09,vandesande13,greene15}. Despite a broad consensus in the literature, there are still major uncertainties with this view of quiescent galaxy formation and evolution: the exact physics of the quenching process is still unknown, and the amount of growth that quiescent galaxies experience after quenching is debated \citep[e.g.][]{carollo13,poggianti13,lilly16}.

The sizes and structures of young quiescent galaxies could hold clues to the physical processes responsible for quenching. Furthermore, comparing the sizes and structures of young and old quiescent galaxies could provide insight into how galaxies grow after quenching. While quiescent galaxies at fixed mass are smaller than their star-forming progenitors \citep[e.g.][]{vanderwel14}, the youngest quiescent galaxies--- or ``post-starburst" galaxies--- appear to be even smaller than their older counterparts \citep{whitaker12_psb,belli15,yano16,almaini17,maltby18,belli19,wu20}. These observations imply that sizes do not simply evolve passively: galaxies appear to shrink when they quench, then grow again as they age.
In addition to their observed small sizes, \psbs are incredibly compact, with high stellar mass densities \citep{almaini17,maltby18}. This may indicate that they reached a quenching density threshold after a rapid core-building process \citep[or ``compaction", e.g.][]{dekel14,zolotov15,tacchella16}.

These results are primarily based on studies of the light profiles of galaxies along the quiescent sequence. However, radial \colorgrads in galaxies are both a potential bias--- they cause half-light and half-mass radii to differ, e.g. \citet{suess19a}--- and a source of additional information, because \colorgrads represent radial variations in the underlying stellar populations of a galaxy.
Studying the evolution of \colorgrads along the quiescent sequence, from post-starburst to old quiescent galaxies, is thus an additional method for probing the physical mechanisms responsible for galaxy growth and quenching.
In this Letter, we present the half-mass radii, \colorgrads, and central mass surface densities of quiescent galaxies as a function of age. 

We assume a cosmology of $\Omega_m=0.3$, $\Omega_\Lambda=0.7$, and $h=0.7$, and a \citet{chabrier03} initial mass function.

%%%%%%%%%%%%%%%%%%%%%%%%%%%%%%%%%%%%%%%%%%%%%
\section{Sample \& Methods}

\begin{figure*}
    \centering
    \includegraphics[width=.9\textwidth]{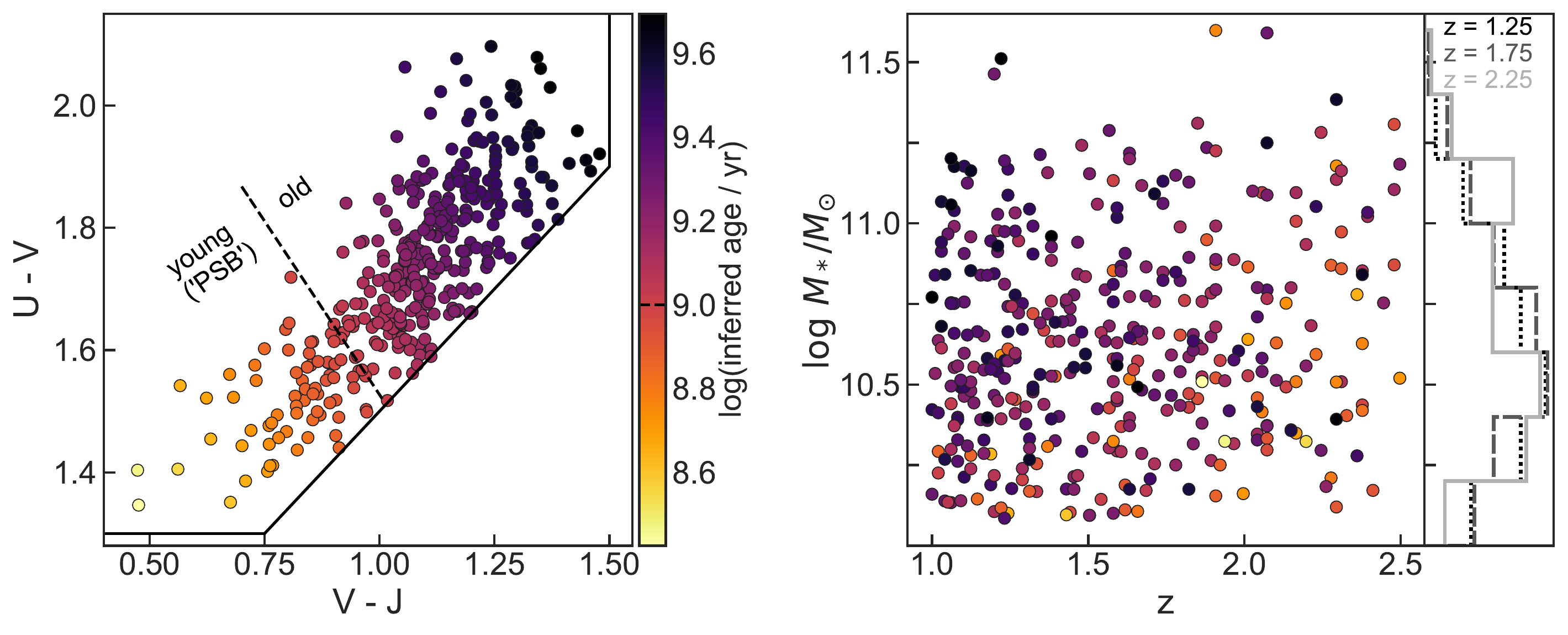}
    \caption{UVJ diagram (left) and stellar mass as a function of redshift (right) for our quiescent galaxy sample. Each galaxy is colored by its inferred stellar age using the \citet{belli19} mapping from UVJ colors to ages. Histograms show the normalized distribution of stellar masses in the three redshift bins used in this work. We only include quiescent galaxies with $M_*>10^{10.1}M_\odot$, where our sample is complete.}
    \label{fig:sample}
\end{figure*}

In this study we use the \citet{suess19a} catalog of \colorgrads and half-mass radii of galaxies at $1.0\le z\le2.5$ in three CANDELS fields \citep{grogin11,koekemoer11}. These measurements were made by fitting spatially-resolved spectral energy distributions (SEDs) to find an observed-frame mass-to-light ratio ($M/L$) gradient. Then, a forward modeling technique was used to account for the point-spread function and recover intrinsic $M/L$ gradients and half-mass radii.

We select 385 quiescent galaxies from the \citet{suess19a} catalog using the \citet{whitaker12_psb} definition for quiescence based on rest-frame UVJ colors. We only include galaxies with $\rm{M}_*>10^{10.1}M_\odot$, where our sample is mass-complete \citep{suess19a}. Our sample lies in the overlap of the CANDELS and ZFOURGE \citep{straatman16} fields, allowing for accurate measurements of both mass profiles and rest-frame SEDs. 

Our study also requires estimates of the age of each galaxy. We use the \citet{belli19} prescription, who use deep continuum spectroscopy of quiescent galaxies to calibrate a mapping between UVJ colors and average stellar ages. Systematic uncertainties on ages calculated using this method are $\sim0.13$~dex. We refer to the youngest quiescent galaxies as ``post-starburst" galaxies; however, see e.g. \citet{wild20} for a discussion of spectroscopically- versus photometrically-selected \psbs.

Figure~\ref{fig:sample} shows our sample in UVJ and mass-redshift space. Each galaxy is colored by its inferred stellar age using the \citet{belli19} technique. 

%%%%%%%%%%%%%%%%%%%%%%%%%%%%%%%%%%%%%%%%%%%%%
\section{Sizes \& color gradients of quiescent galaxies as a function of age}

\begin{figure*}[ht]
    \centering
    \includegraphics[width=.9\textwidth]{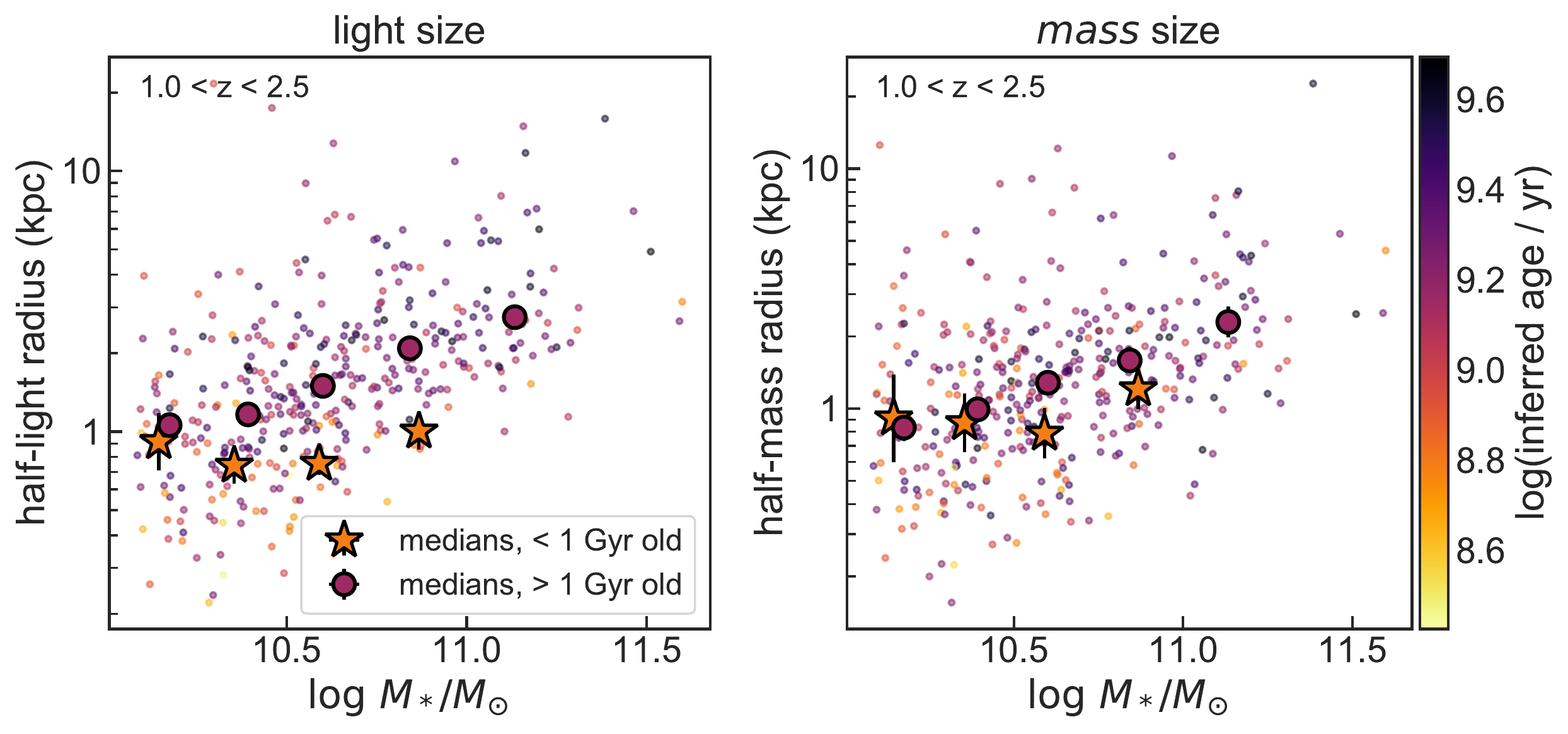}
    \caption{Half-light (left) and half-mass (right) radius as a function of stellar mass for quiescent galaxies at $1.0\le z\le2.5$. Points represent individual galaxies, colored by inferred age. Large purple circles and yellow stars show medians for old (age $>$ 1~Gyr) and young (age $\le$ 1~Gyr) quiescent galaxies; medians are only shown for bins with $\ge5$ galaxies. While there is a 0.3~dex difference in the median half-light radii of young and old quiescent galaxies, this difference is reduced to only $\sim$0.1 dex when considering half-mass radii.}
    \label{fig:mass_size}
\end{figure*}

We begin by examining the size-mass relation for galaxies along the quiescent sequence. The left panel of Figure~\ref{fig:mass_size} shows the size-mass relation using half- light radii at rest-frame 5,000$\AA$ 
\citep[][]{vanderwel14}. Large points show the median sizes of both old (inferred age $>$1~Gyr) and young or `post-starburst' (inferred age $<$1~Gyr) quiescent galaxies in bins of stellar mass. We recover the result that \psbs have smaller half-light radii than older quiescent galaxies \citep{whitaker12_psb,belli15,yano16,almaini17,maltby18,wu18,belli19}. As noted by \citet{almaini17}, this size difference is especially apparent at $M_*>10^{10.5}M_\odot$.

The right panel of Figure~\ref{fig:mass_size} shows the size-mass relation for the same sample, this time using half-{\it mass} radii. The stark difference between the sizes of young and old quiescent galaxies disappears. While young quiescent galaxies have median half-light radii 0.3~dex smaller than their older counterparts, the difference in their median half-mass radii is only $\lesssim$0.1~dex.

We note that the large difference in the half-light radii of young and old quiescent galaxies is partially a redshift effect: the \psbs in our sample have a slightly higher median redshift, and thus smaller half-light radii. However, our general result holds in narrow redshift slices: \psbs have smaller half-light radii than older quiescent galaxies, but their half-mass radii are essentially consistent.

\begin{figure*}
    \centering
    \includegraphics[width=\textwidth]{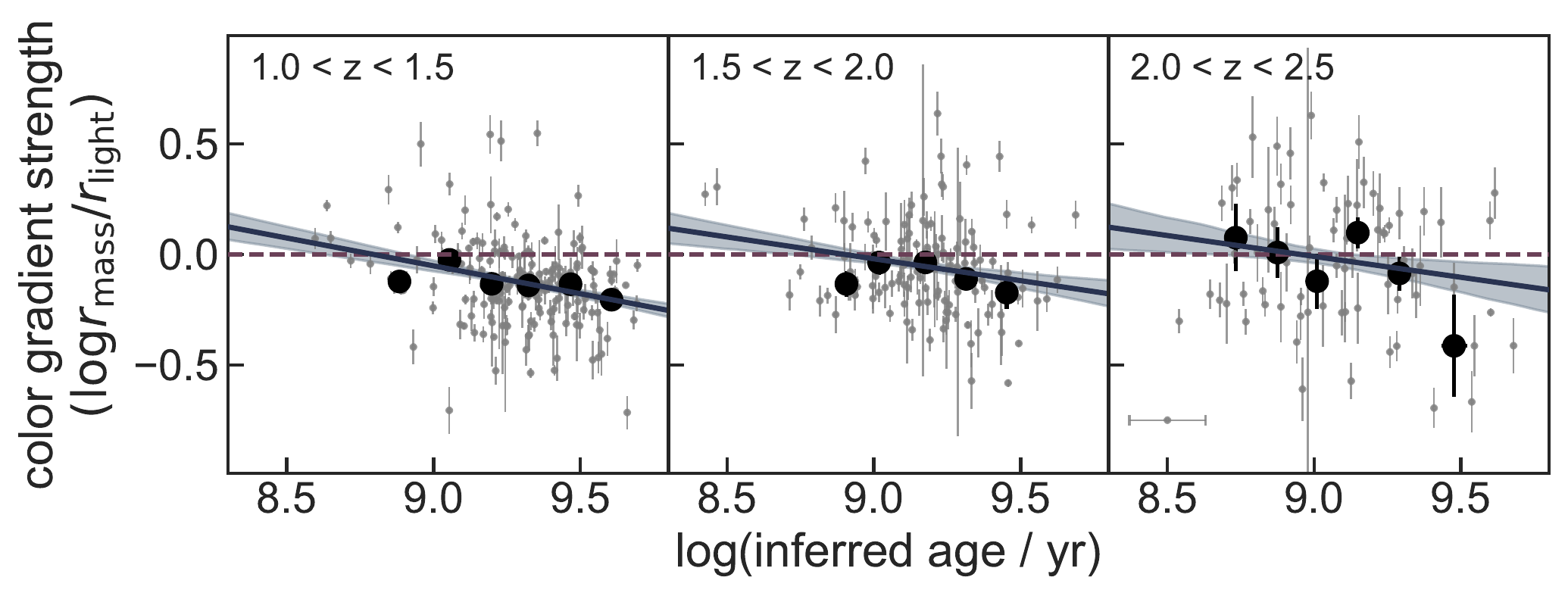}
    \caption{Color gradient strength as a function of inferred age for quiescent galaxies ($M_*\ge10^{10.1}M_\odot$) in three redshift bins. Small grey points represent individual galaxies, black circles are median age bins, and the blue line and shaded region show a best-fit linear relation (and 1$\sigma$ confidence interval, determined by bootstrap resampling) to the individual points. Typical uncertainties in inferred ages are shown by the error bar in the lower left of the right panel. The dashed purple line denotes no color gradient; values above (below) this line indicate bluer (redder) centers. There is a clear trend between \cgstrength and age, such that older galaxies have stronger \colorgrads. The slope of the relation is consistent across redshift. Spearman's $\rho$ (p-values) for each redshift interval are -0.26 ($<0.001$), -0.20 (0.02), and -0.14 (0.25).}
    \label{fig:colorgrad}
\end{figure*}

The fact that the difference between young and old quiescent galaxy sizes significantly shrinks when using half-mass radii is an indication that the two populations have systematically different \colorgrads. 
In Figure~\ref{fig:colorgrad}, we show \cgstrength versus galaxy age. Following \citet{suess19a,suess19b}, we quantify \cgstrength by the log ratio of the half-mass and half-light radii of the galaxy: negative (positive) values indicate that the center of the galaxy is redder (bluer) than the outskirts. We show three redshift slices, because \cgstrength varies with redshift \citep{suess19a,suess19b}. 

At all redshifts, there is a clear relation between \cgstrength and inferred age. A Spearman correlation test indicates that this trend is statistically significant in the two lower-redshift intervals; flatter color gradients and larger error bars on individual galaxies likely contribute to the larger p-value in the highest redshift bin. We fit this \colorgrad-age relation with LEO-Py, including systematic errors on inferred ages and intrinsic scatter around the relation \citep{leopy}. 
We find that young quiescent galaxies have nearly flat \colorgrads. This result agrees with \citet{maltby18}, who found that the optical and near-infrared sizes of \psbs are similar. Galaxies with older stellar ages have increasingly more negative \colorgrads, with redder centers and/or bluer outskirts. The slope of the best-fit relation is consistent across redshift, while the normalization decreases towards lower redshift \citep[consistent with ][]{suess19a}. 

Because \colorgrads tend to be stronger in more massive galaxies \citep[e.g.][]{tortora10,suess19a} and the older galaxies in our sample have higher median stellar masses, we test whether the trend between \cgstrength and galaxy age is a reflection of the color gradient - mass relation. The trend we see in Figure~\ref{fig:colorgrad} persists in mass-matched subsamples, with a consistent slope; this indicates there is a relation between color gradient and age even at fixed mass. We also test whether there is a correlation between residuals in the color gradient-mass relation and residuals in the age-mass relation \citep[e.g.][]{salim15,sanders18}. We find that there is a statistically significant trend between the two sets of residuals at all redshifts, further emphasizing that the trend between \cgstrength and age we see in Figure~\ref{fig:colorgrad} is not purely a mass effect. 

\begin{figure*}
    \centering
    \includegraphics[width=\textwidth]{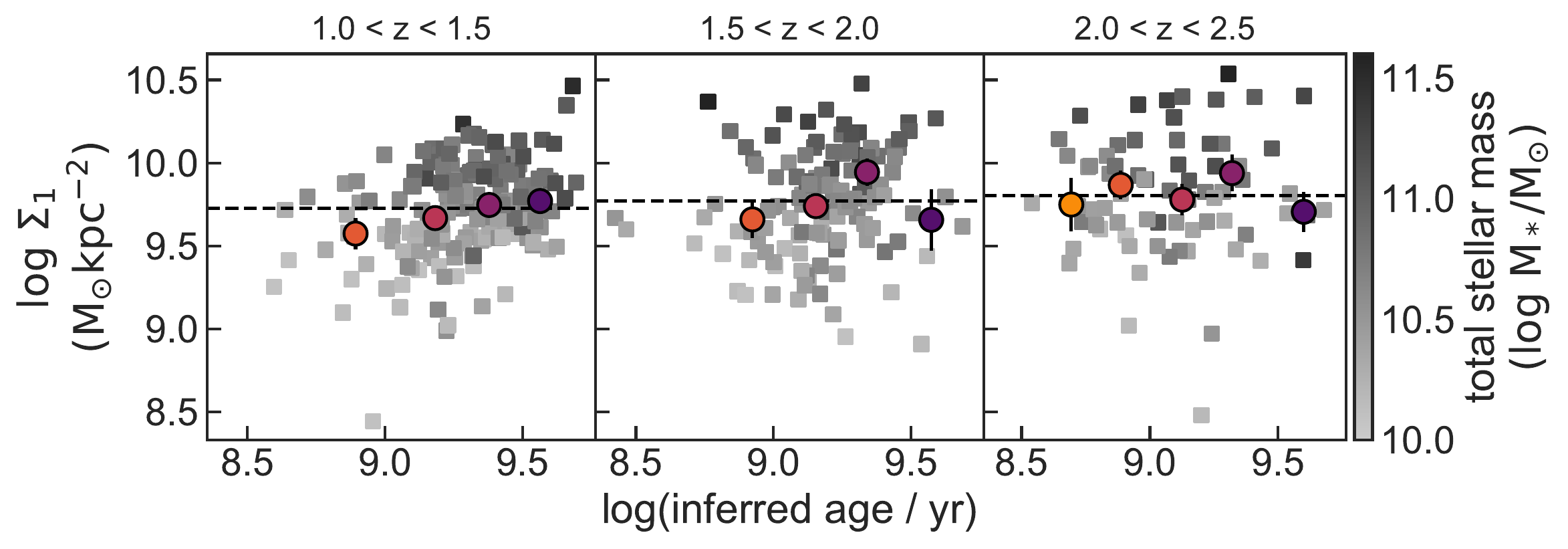}
    \caption{Mass density in the central kiloparsec of the galaxy, $\Sigma_1$, as a function of inferred age in three redshift bins. Squares show individual galaxies, shaded by total stellar mass; colored circles represent median bins in age. The dashed black line shows the median $\Sigma_1$ of all galaxies in a given redshift bin. While $\Sigma_1$ depends on total stellar mass (especially evident in the lowest-redshift bin), it varies only slightly with redshift, and does not depend on stellar age. This supports the idea that galaxies quench with consistently high $\Sigma_1$.}
    \label{fig:sigma}
\end{figure*}

Together, Figures~\ref{fig:mass_size} and \ref{fig:colorgrad} indicate that the observed difference between the half-light radii of young and old quiescent galaxies is {\it not} a true difference in their sizes: instead, it is a systematic difference in their \colorgrads. Older quiescent galaxies have negative \colorgrads, making them appear larger than they truly are; meanwhile, young quiescent galaxies have flat \colorgrads, and appear their `true' size. 
This systematic difference in \cgstrength results in older quiescent galaxies {\it appearing} larger than \psbs, even though their half-mass radii differ by at most $\sim0.1$~dex.

%%%%%%%%%%%%%%%%%%%%%%%%%%%%%%%%%%%%%%%%%%%%%
\section{Central densities of quiescent galaxies as a function of age}

We now turn to the mass profiles of galaxies along the quiescent sequence. Previous studies have found that quiescent galaxies have consistently high central mass densities, implying that the build-up of a dense central core is a prerequisite for quenching \citep[e.g.][]{cheung12,fang13,vandokkum14,tacchella15b,barro17,whitaker17,mosleh17}. We calculate the central mass surface density, $\Sigma_1\equiv \rm{M}_{*,r<{\rm 1kpc}}/\pi$, from the \citet{suess19a} mass profiles. 
In Figure~\ref{fig:sigma} we show $\Sigma_1$ as a function of inferred age. 

While the central mass densities of quiescent galaxies clearly depend on total stellar mass, at fixed mass $\Sigma_1$ is remarkably consistent across age and redshift. We find a best-fit relation\footnote{This least-squares fit was performed using the python \texttt{lmfit} package for consistency with \citet{fang13} and \citet{barro17} in order to facilitate a direct comparison of the results. A LEO-Py fit including systematic error bars on the independent variables as well as intrinsic scatter around the relation finds a slightly steeper mass dependence and slightly shallower redshift dependence, consistent with the \texttt{lmfit} results within 1.5$\sigma$.} of 
\begin{equation}
    \begin{split}
        \log{\Sigma_1} = & (9.40 \pm 0.06) + (0.69 \pm 0.03) (\log{M_*/M_\odot} - 10.5) \\ & + (0.78 \pm 0.17) \log{(1+z)} \\ &+ (-0.03 \pm 0.04) (\log{\rm{age/yr}}-9.5)
    \end{split}
    \label{eqn:fit}
\end{equation}
Previous studies have found a dependence of $\Sigma_1\propto (M_*/M_\odot)^{0.64}$ \citep{fang13,barro17,saracco17,tacchella17}, consistent with our fit within 1$\sigma$ error bars. Both \citet{barro17} and \citet{mosleh17} calculated slightly shallower redshift evolution, $\Sigma_1\propto(1+z)^{0.55-0.68}$, but again the values are consistent within 1$\sigma$ errors. We note that the redshift evolution we find is still relatively slow, and indicates that $\Sigma_1$ decreases by only $\sim0.13$~dex between our highest and lowest redshift bin.

We find that the dependence of $\Sigma_1$ on galaxy age is consistent with zero (Equation~\ref{eqn:fit}, Figure~\ref{fig:sigma}). 
This implies that more compact galaxies do {\it not} quench earlier: in that case, older galaxies at fixed mass would have higher $\Sigma_1$. We note that the most massive galaxies quench first \citep[e.g.,][]{cowie96}, and more massive galaxies tend to have higher $\Sigma_1$. This effect fully accounts for the slight trend of increasing $\Sigma_1$ in older and more massive galaxies in our $1.0\le z\le1.5$ bin, and suggests that the results of \citet{saracco17}--- who find that older and more massive galaxies have higher $\Sigma_1$--- are driven by mass, not age, effects.  

While we find that $\Sigma_1$ depends on stellar mass and redshift but not age, \citet{estrada20} find that $\Sigma_1$ depends on stellar mass and formation redshift. These two interpretations are not inconsistent: $\Sigma_1$ is highest in massive galaxies that form early in the universe, but does not appear to evolve as galaxies age after quenching.

\citet{tacchella17} find that older galaxies at fixed mass have higher $\Sigma_1$, inconsistent with our results. However, their study was performed at $z=0.05$: this discrepancy may reflect differences in quenching mechanisms or the properties of recently-quenched galaxies at $z>1$ and $z\sim0$ \citep[e.g.][]{maltby18}.

%%%%%%%%%%%%%%%%%%%%%%%%%%%%%%%%%%%%%%%%%%%%%
\section{Discussion}

\begin{figure*}
    \centering
    \includegraphics[width=.95\textwidth]{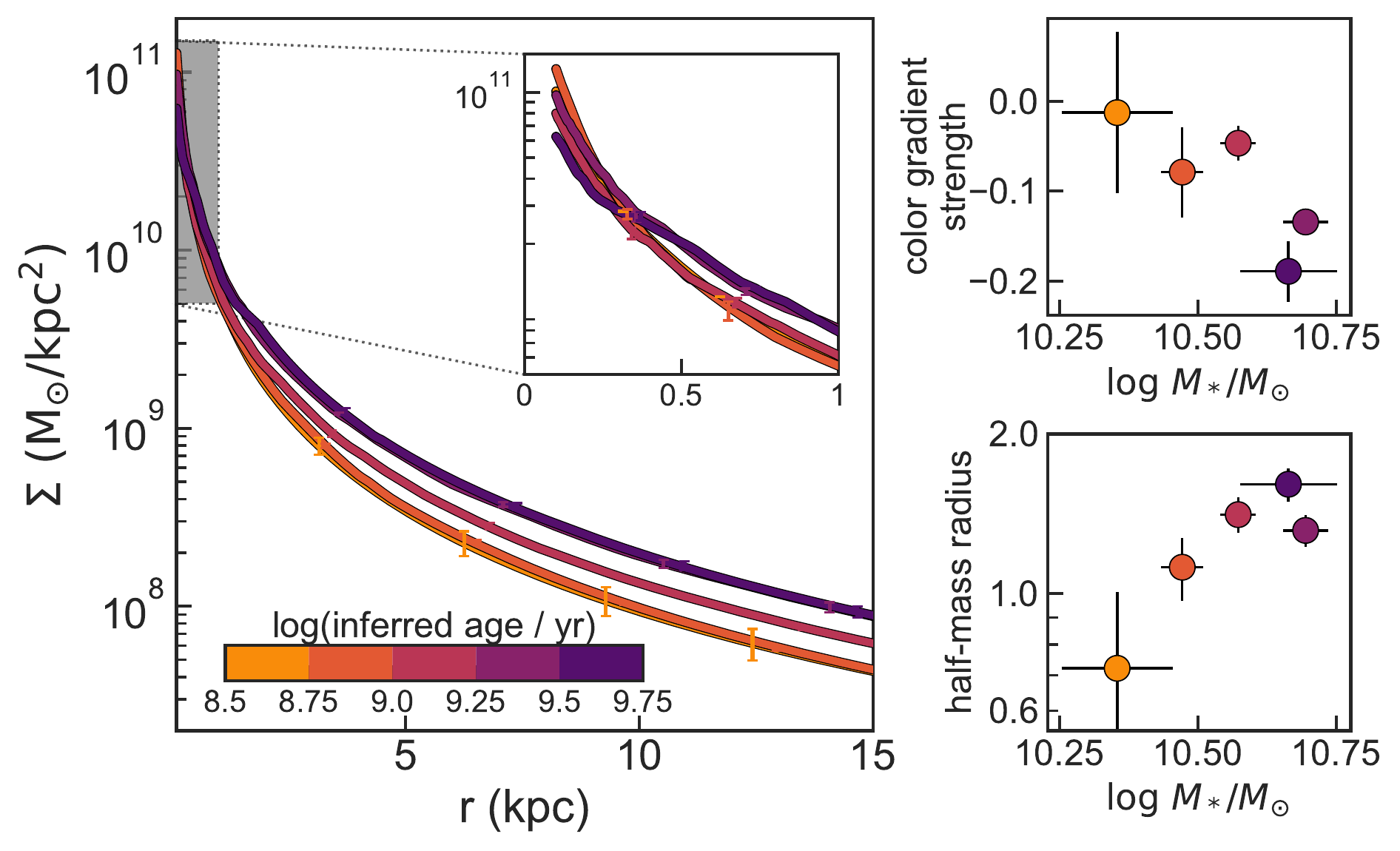}
    \caption{Left: mass profiles of quiescent galaxies ($1.0\le z\le2.5$) split into age bins. The central mass profiles are similar across age; extra mass in older galaxies is deposited mostly in the wings. Right: median \cgstrength ($\log{r_{\rm{mass}}/r_{\rm{light}}}$, top) and half-mass radius (bottom) as a function of median stellar mass for the same age bins. \colorgrads are flatter in younger galaxies; the older galaxies in our sample are also generally more massive (and thus larger).}
    \label{fig:profiles} 
\end{figure*}

In this Letter we investigate the half-mass radii, \colorgrads, and central mass densities of $1.0\le z\le2.5$ quiescent galaxies as a function of age. We find that \psbs are {\it not} significantly smaller than older quiescent galaxies of the same stellar mass; they only appear smaller because they have systematically flatter \colorgrads (Figures \ref{fig:mass_size} \& \ref{fig:colorgrad}). At the same time, we see that \psbs are compact, with central mass densities consistent with those of older quiescent galaxies of the same mass (Figure \ref{fig:sigma}). 

These observations can help address two separate questions: what mechanism caused \psbs to stop forming stars? And how do quiescent galaxies evolve after they quench? To place our observations in context, Figure~\ref{fig:profiles} shows the median mass profiles, \colorgrads, and half-mass radii of all quiescent galaxies in our sample, binned by age. Each binned point includes galaxies across our full mass and redshift range.
We bin across redshift because neither half-mass radii nor $\Sigma_1$ evolve significantly over our $1.0\le z\le2.5$ range \citep[Eqn \ref{eqn:fit}; ][]{suess19b}. However, we note that the older quiescent galaxies in our sample are also more massive; therefore, each age bin probes a different stellar mass. This is not simply selection bias: our sample is complete to the same mass limit in each redshift bin (Figure~\ref{fig:sample}).   

First, we consider how \psbs halt their star formation. If slow or gradual processes such as gas exhaustion are responsible, we would expect the sizes, structures, and \colorgrads of \psbs to resemble those of their progenitors. 
However, we see in Figures~\ref{fig:colorgrad} and \ref{fig:profiles} that \psbs have flat \colorgrads, while \citet{suess19a} found that star-forming galaxies at similar masses and redshifts typically have negative \colorgrads. Because \colorgrads indicate radial variations in the underlying stellar populations, in order to alter the \colorgrads in a galaxy some physical process must either create new stellar populations or re-arrange existing ones. Candidate processes include mergers--- which could flatten or destroy radial \colorgrads--- or a central starburst, which could create an excess of young stars at the center of the galaxy and flatten a pre-existing negative color gradient.  
Both of these processes are popular candidates for the fast quenching process, whereby galaxies rapidly build a massive core before shutting off their star formation \citep[e.g.,][]{zolotov15,tacchella15b,barro17,woo19}. 

Second, we turn to the question of how galaxies evolve after they quench. In Figures~\ref{fig:colorgrad} and \ref{fig:profiles} we see that older and more massive galaxies also have stronger negative \colorgrads; this indicates that some process must re-establish negative \colorgrads after quenching. At the same time, we see that the shapes of their mass profiles change (left, Figure~\ref{fig:profiles}) and their half-mass radii grow (bottom right, Figure~\ref{fig:profiles}). 
Older quiescent galaxies have slightly less centrally-peaked mass profiles, potentially due to post-quenching adiabatic expansion \citep[e.g.,][]{choi18} or dynamical friction from mergers \citep[e.g.,][]{naab09}. Older quiescent galaxies also have more mass at large radii, resulting in larger average sizes.  
These observations are consistent with a picture where quiescent galaxies grow ``inside-out" at late times: minor mergers deposit younger and/or lower-metallicity stars at the outskirts of galaxies, causing negative radial \colorgrads \citep[e.g.,][]{bezanson09,naab09,suess19a}. In this scenario, we would expect older quiescent galaxies--- which have had time for more inside-out growth--- to have more negative \colorgrads, higher total stellar masses, more mass in their outskirts, and thus larger sizes. This is a good match for what we see in Figure~\ref{fig:profiles}.

This picture relies on interpreting the ages of quiescent galaxies as an evolutionary sequence. It is important to verify this assumption and ensure that our results are not driven by selection effects.  
One possibility is that the trends we see are caused by evolution in the properties of recently-quenched galaxies \citep[``progenitor bias", e.g.][]{carollo13,poggianti13}. In the \citet{lilly16} model, star-forming galaxies have negative \colorgrads, which strengthen as galaxies grow; these \colorgrads cease to evolve after quenching. In a pure progenitor bias scenario, older quiescent galaxies formed earlier and should thus have flatter \colorgrads, because star-forming galaxies at higher redshift have flatter \colorgrads \citep{suess19a,suess19b}. However, we see that older quiescent galaxies actually have {\it steeper} \colorgrads (Figure~\ref{fig:colorgrad}). This model is thus inconsistent with our observations. Furthermore, while quiescent galaxies at high redshift have flat or even positive \colorgrads, old quiescent galaxies at low redshift have negative \colorgrads (Figure~\ref{fig:colorgrad}). This also implies that \colorgrads evolve after quenching, inconsistent with the \citet{lilly16} model. Finally, the observed evolution of half-mass radii at $1.0\le z\le2.5$ can be almost entirely explained by minor mergers alone, without the need for progenitor bias \citep{suess19b}. 

A second possibility is that not all quiescent galaxies go through a post-starburst phase, so not all of the progenitors of old quiescent galaxies are included in our sample. Indeed, the number densities of \psbs are not sufficiently high to explain the full growth of the quiescent sequence at these redshifts \citep[e.g.,][]{wild16,belli19}. Recent work has suggested that green valley galaxies--- which lie just outside of our UVJ selection--- may represent a second, slower, path to join the quiescent sequence \citep[e.g.,][]{bremer18,wu18,belli19,woo19}. If green valley galaxies quench without significantly altering their structural properties \citep{wu18}, we would expect them to have negative \colorgrads inherited from their star-forming progenitors. The trend we see between \cgstrength and age could thus be explained without appealing to post-quenching evolution {\it if} all old quiescent galaxies with negative \colorgrads quenched via the green valley, and all old quiescent galaxies with flat \colorgrads quenched via the post-starburst route. However, the presence of old quiescent galaxies with negative \colorgrads in our highest-redshift bin--- where quenching through the green valley is less prevalent \citep[e.g.,][]{belli19}--- indicates that this selection effect does not fully account for the trends we see in this Letter. We will investigate the \colorgrads and half-mass radii of green valley galaxies in detail in a forthcoming paper.

Altogether, our observations appear consistent with a picture where \psbs quench after experiencing a rapid core-building process; after quenching, we observe changes in the quiescent population that appear consistent with inside-out growth via minor mergers. This work was enabled by studying radial \colorgrads and their evolution directly: not only are \colorgrads a valuable observable in their own right, they are essential for obtaining an unbiased view of galaxy size evolution. In the future, when upcoming missions such as {\it JWST} provide spatially-resolved information through the infrared, we hope that this work will spur future investigations into \colorgrads as a powerful probe of galaxy evolution. 

\acknowledgements 
KAS thanks Sirio Belli, Vince Estrada-Carpenter, and Sandro Tacchella for inspiring discussions and helpful comments. We also thank the anonymous referee whose suggestions improved this paper.
This work is funded by grant AR-12847, provided by NASA though a grant from the Space Telescope Science Institute (STScI) and by NASA grant NNX14AR86G. This material is based upon work supported by the National Science Foundation Graduate Research Fellowship Program under grant No. DGE 1106400.

\bibliographystyle{aasjournal}
\bibliography{psb_bib}

\end{document}